\documentclass[11pt]{article}

\usepackage{amsmath}
\usepackage{graphicx}
\usepackage{geometry}
\usepackage{setspace}
\usepackage{natbib}
\title{A novel two-stage parameter estimation framework integrating Approximate Bayesian Computation and Machine Learning: The ABC-RF-rejection algorithm}
\author{Renata Retkute\,$^{1,}$\thanks{Corresponding author: rr614@cam.ac.uk}\,$^{ }$ and Christopher A. Gilligan\,$^{1}$\\
$^{1}$Epidemiology and Modelling Group, Department of Plant Sciences,\\  University of Cambridge, UK }
\date{\today}

\begin{document}

\maketitle

\begin{abstract}
We introduce a novel two-stage parameter estimation framework designed to improve computational efficiency in settings involving complex, stochastic, or analytically intractable dynamic models. The proposed method, termed \textit{ABC-RF-rejection}, integrates Approximate Bayesian Computation (ABC) rejection sampling with Random Forest (RF) classification to efficiently screen parameter sets that produce simulations consistent with observed data. We evaluate the performance of this approach using both a deterministic Susceptible-Infected-Removed (SIR) epidemic model and a spatially explicit stochastic epidemic model. Results indicate that \textit{ABC-RF-rejection} achieves substantial gains in computational efficiency while maintaining parameter inference accuracy comparable with standard ABC rejection methods. Finally, we apply the algorithm to estimate parameters governing the spatial spread of cassava brown streak disease (CBSD) in Nakasongola district, Uganda.
\end{abstract}

\section{Introduction}
Parameter inference for complex models, particularly those characterised by stochasticity, spatial structure, or analytically intractable likelihoods, remains a significant challenge in quantitative epidemiology and ecological modelling \citep{li2025}. Approximate Bayesian Computation (ABC) provides a solution by approximating posterior distributions through simulations and comparison of summary statistics \citep{Minter2019}. However, ABC rejection methods can be computationally inefficient, especially when acceptance rates are low. To address this limitation, we propose a novel hybrid framework, \textit{ABC-RF-rejection}, which integrates ABC rejection sampling with a supervised machine learning classifier to improve the efficiency of identifying parameter sets likely to yield acceptable simulations when compared with observational data. Building upon ensemble learning principles \citep{Breiman2001}, our approach leverages Random Forest classifiers to predict acceptance probabilities, thus focusing computational resources on high-probability regions of parameter space.

 We evaluate the performance of the ABC-RF-rejection with three case studies. Case study 1 is an introductory example that illustrates the application of the ABC-RF-rejection algorithm to a deterministic epidemic model, with a particular focus on improvements in computational efficiency in comparison with a classical ABC-rejection algorithm. Case study 2 is a spatially explicit stochastic epidemic model \citep{Retkute2025}, where we generate synthetic data with known parameters and seek to recover the parameters using ABC-RF-rejection. To address the challenge of limited surveillance data, we introduce and validate summary statistics specifically designed for such conditions. Finally, in case study 3 we use the algorithm to estimate parameters for cassava brown streak virus (CBSV) spread in Nakasongola district in central Uganda.  Cassava is the second most important source of calories in sub-Saharan Africa (SSA) \citep{Adebayo2023}. Cassava brown streak disease (CBSD), caused by \textit{Ipomoviruses}from the family Potyviridae, poses a major threat to cassava production across the region \citep{Alicai2007}. The virus isspread by  the whitefly vector \textit{Bemisia tabaci} and by farmers exchanging virus-infected planting material \citep{Maruthi2005}. We chose Nakasongola district as it is an exemplary setting for studying CBSV dynamics due to its high disease prevalence and widespread cultivation of cassava by  smallholder farmers \citep{McQuaid2015, McQuaid2017}. 

Our results demonstrate that the ABC-RF-rejection algorithm achieves substantial improvements in computational efficiency while preserving parameter inference accuracy comparable with that of conventional ABC rejection techniques. Our approach offers an adaptable and computationally efficient solution for parameter estimation in complex disease models, particularly in settings with limited data, spatial heterogeneity, and the need for rapid, evidence-based decision-making. 
 
\section{Overview of the ABC-RF-Rejection algorithm}
We present a novel two-stage parameter estimation framework designed to enhance sampling efficiency in complex modelling scenarios. This approach, which we term ABC-RF-rejection, integrates Approximate Bayesian Computation (ABC) rejection with Random Forest (RF) classification to selectively identify parameter sets likely to yield simulations consistent with observed data.

In the first stage, we apply the ABC rejection algorithm. Approximate Bayesian Computation methods approximate the posterior distribution of model parameters in situations where the likelihood is analytically intractable or the available data are noisy, coarse, or high-dimensional \citep{Minter2019}. The ABC rejection approach accepts parameter sets - referred to as particles - when the distance between summary statistics derived from simulated and observed data falls below a user-defined threshold. Parameters are sampled from the prior distribution, simulations are conducted, and summary statistics are computed. Each particle is assigned a status of accepted (label = 1) or rejected (label = 0) based on the specified threshold criterion.

In the second stage, we employ a supervised machine learning model to learn the relationship between parameter values and the likelihood of acceptance as determined in the first stage. Specifically, we use a Random Forest classifier, a robust ensemble method that aggregates predictions from multiple decision trees to improve generalization and reduce variance \citep{Breiman2001}. The Random Forest model is trained using the labelled dataset generated by the ABC rejection step. Classification is implemented using the \texttt{randomForest}  package in R.  After model training, we generate a substantially larger set of candidate particles. The trained Random Forest model is then used to predict the probability of acceptance for each new particle. Only those particles with high predicted acceptance probabilities are used for further simulation and posterior approximation.

\section{Case studies}

\subsection{Application to a deterministic SIR Model}
\subsubsection{Model and data}
The deterministic SIR (Susceptible-Infectious-Recovered) model is a foundational framework in infectious disease epidemiology used to describe the spread of contagious diseases within a closed population \citep{Bjrnstad2020}. It divides the population into three compartments: susceptible individuals (S) who can contract the disease, infectious individuals (I) who can transmit the disease, and recovered individuals (R) who have gained immunity. The model is governed by a set of ordinary differential equations that describe the rates at which individuals transition between compartments based on two key parameters: the transmission rate ($\beta$) and the recovery rate ($\gamma$). The dynamics of each group can be expressed through a system of three interconnected differential equations:
\begin{align}
\frac{dS}{dt} &= -\beta \frac{SI}{N}, \\
\frac{dI}{dt} &= \beta \frac{SI}{N} - \gamma I, \\
\frac{dR}{dt} &= \gamma I,
\end{align}
Here  $N = S + I + R$ is the total population size.

The deterministic nature of the SIR model means it produces the same output for a given set of parameters and initial conditions, without incorporating randomness or stochastic variation.   We simulated an epidemic using $\beta = 1.5$, $\gamma = 0.5$, $N = 1000$, and initial conditions $I_0 = 1$ over a 20-day period, recording infected individuals at $t = 1, 5, 9, 13, 17$ days (Fig. \ref{figCase1}A).

\subsection{Algorithm implementation and results}
\begin{figure}[t]
\includegraphics[width=1.0\textwidth]{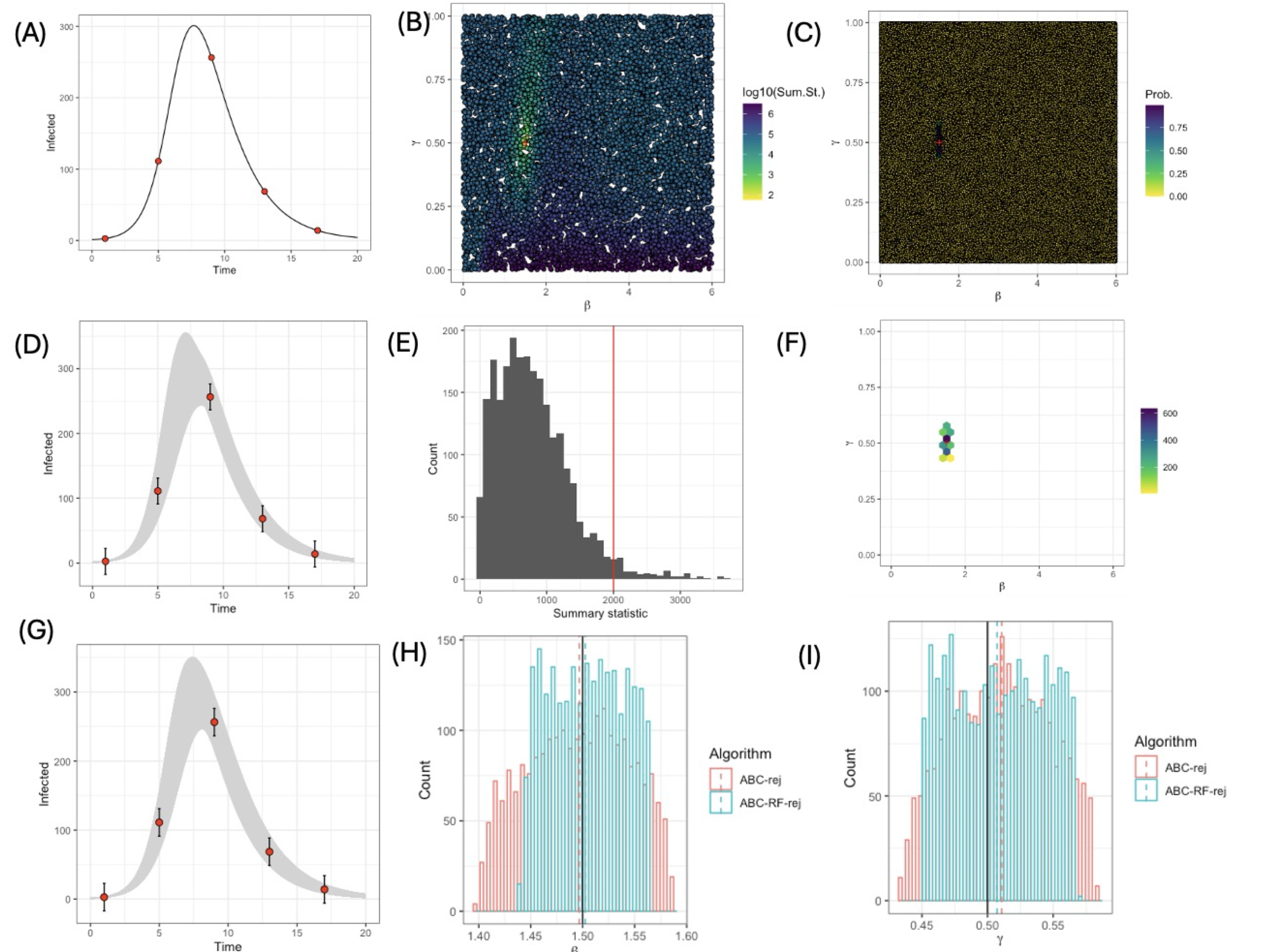}
\caption{(A) Simulated epidemic based on SIR model (black curve) and observed data (red dots).  (B) Sampled $10^5$ particles after Stage 1, colours indicate log10 transformed values of summary statistics. True parameter values shown as a red star. (C) Sampled $10^6$ particles and probability of acceptance based on the trained RF model. (D) Posterior simulations for ABC-RF-rejection (grey curves) and observed data (red dots). Error bars are equal to $\pm20$ individuals.  (E) Distribution of summary statistics for particles with probability of acceptance more than 0.75. Vertical red line shows threshold of summary statistic. (F) Distribution of accepted particles based on ABC-rejection algorithm. (G) Posterior simulations for ABC-rejection (grey curves) and observed data (red dots). Error bars are equal to ±20 individuals. (H)-(I) Posterior distribution of transmission and recovery rates. Dashed lines show the mean, and black line is the true value of the parameter.}\label{figCase1}
\end{figure}

As a key component of the algorithm, we defined the summary statistic as the sum of differences between observed and simulated values:
\begin{equation}
ss_{\text{obs, sim}} = \sum_{i=1}^n \left( I_{t_i}^{\text{obs}} - I_{t_i}^{\text{sim}} \right)^2,
\end{equation}
where $n$ is the number of observations. We set the threshold $ss^{th}=5 \times 20^2=2000$, corresponding to a tolerance of 20 individuals per observation.
In the first stage of the ABC-RF-rejection algorithm, we sampled $10^5$ particles from the prior, ran model simulations, and computed the summary statistics (Fig. \ref{figCase1}B). Particles with $ss < 2000$ were labelled as accepted (1), others as rejected (0) and used to train a Random Forest classifier. 

The prior distributions for the parameters were defined as follows: $\beta\sim U[0,6]$,and $\gamma \sim U[0,1]$. In the second stage, we sampled $10^6$ new particles, predicted acceptance probabilities using the trained model (Fig. \ref{figCase1}C), and retained particles with predicted probability $\geq0.75$ as approximate posterior samples. Simulations based on these particles are shown in Figure \ref{figCase1}D). Of the 2,368 selected particles, over 96\% had $ss<2000$ (Fig. \ref{figCase1}E), demonstrating high efficiency, with potential for further improvement by increasing the probability threshold.

To benchmark computational gains, we compared the ABC-RF-rejection with standard ABC rejection. To obtain the same number of accepted particles (2,368), standard ABC-rejection required 666,972 simulations, yielding an acceptance rate equal to 0.0035. Posterior simulations and parameter distributions are shown in Figures \ref{figCase1}F-G. While both methods produced posterior means close to the true parameter values, ABC-RF-rejection resulted in notably narrower posterior distributions, indicating greater precision (Figures \ref{figCase1}H-I).

\subsection{Application to a stochastic spatial epidemic model}

\subsubsection{Model and data} \label{subs:stochspatmod}

We evaluate the performance of our parameter inference algorithm and assess the effectiveness of the proposed summary statistics within the context of a spatially explicit stochastic model. We use the model proposed in \citep{Retkute2025}. To model the spatiotemporal dynamics of a pathogen, we discretize the landscape into a regular grid of cells. Each grid cell $i$ is characterized by a host density $h_i)$, representing the proportion of land area under crop cultivation. Grid cells are assigned one of two epidemiological states: susceptible (S) or infected (I). A grid cell is considered susceptible if all crop plants within it are healthy; the infection of the first plant causes the entire grid cell to transition to the infected state.

New infections in susceptible cells may arise via primary transmission, representing external introductions (e.g., long-distance movement of infected planting material), or through secondary transmission from neighbouring infectious grid cells.

After a grid cell becomes infected, the local progression of infection follows a deterministic logistic growth model that describes the temporal increase in within-cell infection. For a grid cell $i$ infected at time $t$, the proportion of infected hosts at time $t$ is given by:
\begin{equation}\label{eq:localgrowth}
    \rho_i(t) = \frac{1}{1+\left(\frac{1}{p_0}-1 \right) e^{-r(t-t_i)}},
\end{equation}
where $p_0$ is the initial infection prevalence when the cell transitions to the infected state, and $t$ is the intrinsic rate of local epidemic growth. 

The instantaneous hazard $\lambda_i(t)$ of infection for a susceptible cell $i$ at time $t$ is defined as:
\begin{equation}
    \lambda_i(t)= h_i  \left(\epsilon + \beta \sum_{j \in I_t, j \neq i} h_j \rho_j(t) K(\alpha, d_{i,j})\right),
\end{equation}
where $\epsilon$ is the primary transmission rate, $\beta$ is the secondary transmission rate; $I_t$, is an index set of infectious grid cells at time $t$; $d_{i,j}$ is the Euclidean distance between the centres of grid cells $i$ and $j$.

We adopt an exponential dispersal kernel of the form:
\begin{eqnarray}
K(\alpha, d)=\frac{1}{2\pi \alpha} \exp(- d/\alpha).
\end{eqnarray}
Here $\alpha$ represents the dispersal scale. 

To account for the stochasticity of infection events under dynamic exposure rates, we simulate the transmission process using the Gillespie stochastic simulation algorithm \citep{Gillespie1976, Gillespie1977}.

 We considered a host to be on a $100 \times 100$ grid,  with 0.75 occupancy at each grid cell. The true parameter values were set to $\epsilon = 0.0001$, $\beta = 8$, and $\alpha = 0.5$. The epidemic was initiated from a single infected cell at the landscape centre and simulated over a four-year period. The “observed” outbreak, serving as a synthetic target for inference, is shown in Figure \ref{figCase2}A. The outbreak resulted in 320 infected cells (Figure \ref{figCase2}B).

\subsubsection{Algorithm implementation and results}
Two summary statistics were employed for parameter estimation. The first statistic, $ss_1$, quantified the annual radial spread of the epidemic, defined as the maximum Euclidean distance between the initial infection and all infected cells by the end of each year ($t = 1, 2, 3, 4$). This statistic was computed as the sum of squared deviations between simulated and observed maximum distances:
\begin{eqnarray}
    ss_2=\sum_{t=1}^4\left(d_t^{obs}-d_t^{sim}  \right)^2.
\end{eqnarray}
To constrain the acceptable range of deviation, we imposed a threshold whereby the maximum distance could differ by no more than 5km per year, resulting in a total allowable discrepancy of $ss_1^{th}=4\times5^2=100$.

\begin{figure}[t]
\includegraphics[width=1.0\textwidth]{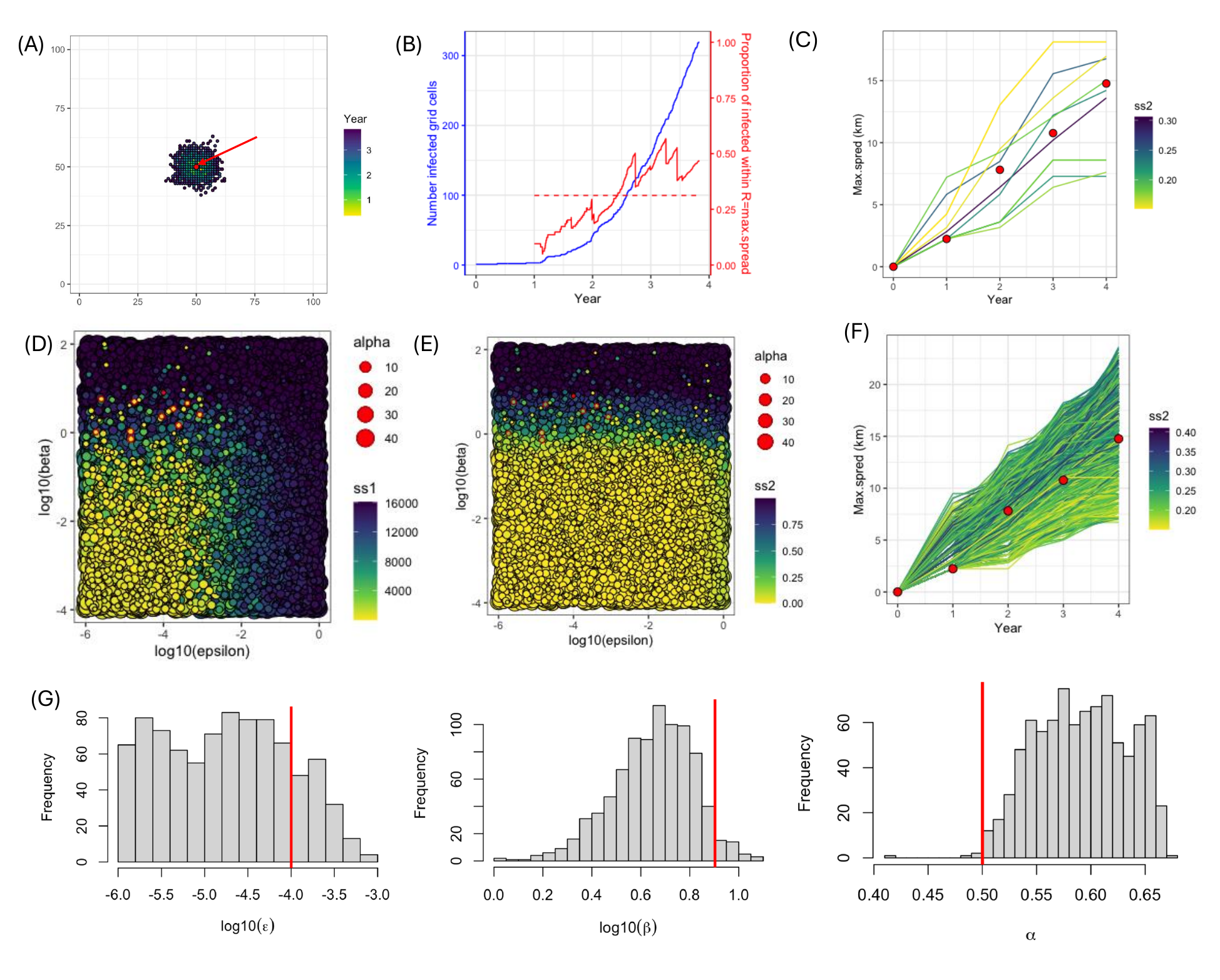}
\caption{(A) Spatial distribution of the simulated outbreak. (B) Temporal changes in number of infected grid cells and proportion of infected grid cells within maximum spread radius. (C) Summary statistics of particles accepted after stage 1. (D) Sampled parameters and values of $ss_1$. Red circles show accepted particles and the full circle is the true values parameter. (E) Sampled parameters and values of $ss_2$. Red circles show accepted particles and the full circle is the true values parameter. (F) Summary statistics of particles accepted after the stage 2. (G) Marginal posterior distributions of parameters. Red lines show the true values of the parameters.}\label{figCase2}
\end{figure}

We did not use the total number of infected grid cells as a summary statistic due to practical limitations: in realistic surveillance scenarios, full grid coverage is rarely achieved. Furthermore, given the model’s stochastic nature, the number of infected cells varies significantly between simulations, even under identical parameter settings. Instead, our second summary statistic captured the relative intensity of infection. This was approximated as the ratio of infected cells to the total number of cells within a circular area centred at the initial infection site, with a radius equal to the maximum spread at time $t$. The red curve in Figure \ref{figCase2}B depicts the temporal trajectory of this ratio, and the dashed line the average. The second summary statistic, denoted $ss_2$, was defined as simulated proportions at the end of year four, which we required to be close to the observed proportion, which we set equal to 0.28 (red dashed line in Fig. \ref{figCase2}B).

The prior distributions for the parameters were defined as follows: $\log10(\epsilon)\sim U[-6,0]$, $\log10(\beta)\sim U[-4,2]$,and $\alpha \sim U[0.01,50]$. In the first stage of the ABC-RF-rejection algorithm, $10^5$ parameter sets were sampled from the prior distributions. Model simulations were performed for each set, and corresponding summary statistics were computed. Only nine parameter sets met the acceptance criteria based on the defined thresholds (Fig. \ref{figCase2}C). Distributions of the summary statistics across all particles are presented in Figures \ref{figCase2}D–E, with accepted particles highlighted in red and the true parameter set indicated by a filled red circle.

In the second stage, the accepted parameter sets were labelled as “accepted” (class 1) and used to train a Random Forest classifier. Subsequently, $10^7$ parameter sets were sampled from the priors, and the trained model was used to estimate their probability of acceptance. Parameter sets with predicted acceptance probabilities exceeding 0.5 were selected for further simulation, using a lower threshold than in previous sections to account for increased stochastic variability. Ultimately, 867 out of 4,473 parameter sets satisfied both summary statistic criteria and formed the empirical posterior distribution. Marginal posterior distributions are shown in Figure \ref{figCase2}F, and we recovered true parameter values, as these fall within the inferred posteriors.

\subsection{Application to cassava brown streak disease (CBSD) spread}
\subsubsection{Model and data}
We employed the stochastic, spatially explicit model described in Section \ref{subs:stochspatmod}, which accounts for both within-cell and between-cell transmission of CBSV across a gridded landscape. The dynamics of CBSV spread within an infected grid cell were parameterised using experimental data from controlled field trials conducted at two sites in Uganda \citep{Katono2015}. We fitted the logistic function to the empirical data using a non-linear least squares approach. Fitted logistic curve and experimental data are shown in Fig. \ref{figCase1}A. The estimated values are $r=13.2$ year$^{-1}$, and $p_0=0.004$.

Shapefiles for Uganda's second-level administrative boundaries were sourced from the Database of Global Administrative Areas \citep{GADM2018}. Spatial information on cassava cultivation was obtained from the CassavaMap \citep{Szyniszewska2020}, which provides estimates of cassava harvested area at 1 km resolution. We converted harvested area values into a fractional host density ($h_i$) representing the proportion of land within each model grid cell occupied by cassava (Fig.\ref{figCase1}B). Surveillance data on CBSD spread in Uganda came from \citep{Alicai2019}. We extracted data for Nakasongola district from 2008 to 2017, annual presence and absence reports are shown in Figure \ref{figCase1}C. There was no surveillance in 2016 \citep{Alicai2019}. We assumed that the outbreak started at the location of a single positive report in 2008 in the west part of Nakasongola district.

\subsubsection{Algorithm implementation and results}
We used two summary statistics for parameter estimation:  $ss_1$, to quantify the annual radial spread of the epidemic; and $ss_2$, to characterize strength of infection and representing the observed infection proportion. We calculated a threshold for $ss_1$ by requiring that the maximum distance differed by no more than 10 km per year. As only a subset of grid cells was surveyed, these proportions could not be directly used. Instead, they informed target values for the second summary statistic, which captured outbreak intensity. We set a target infection proportion after nine years to $0.15 \pm 0.1$.

\begin{figure}[t]
\includegraphics[width=1.0\textwidth]{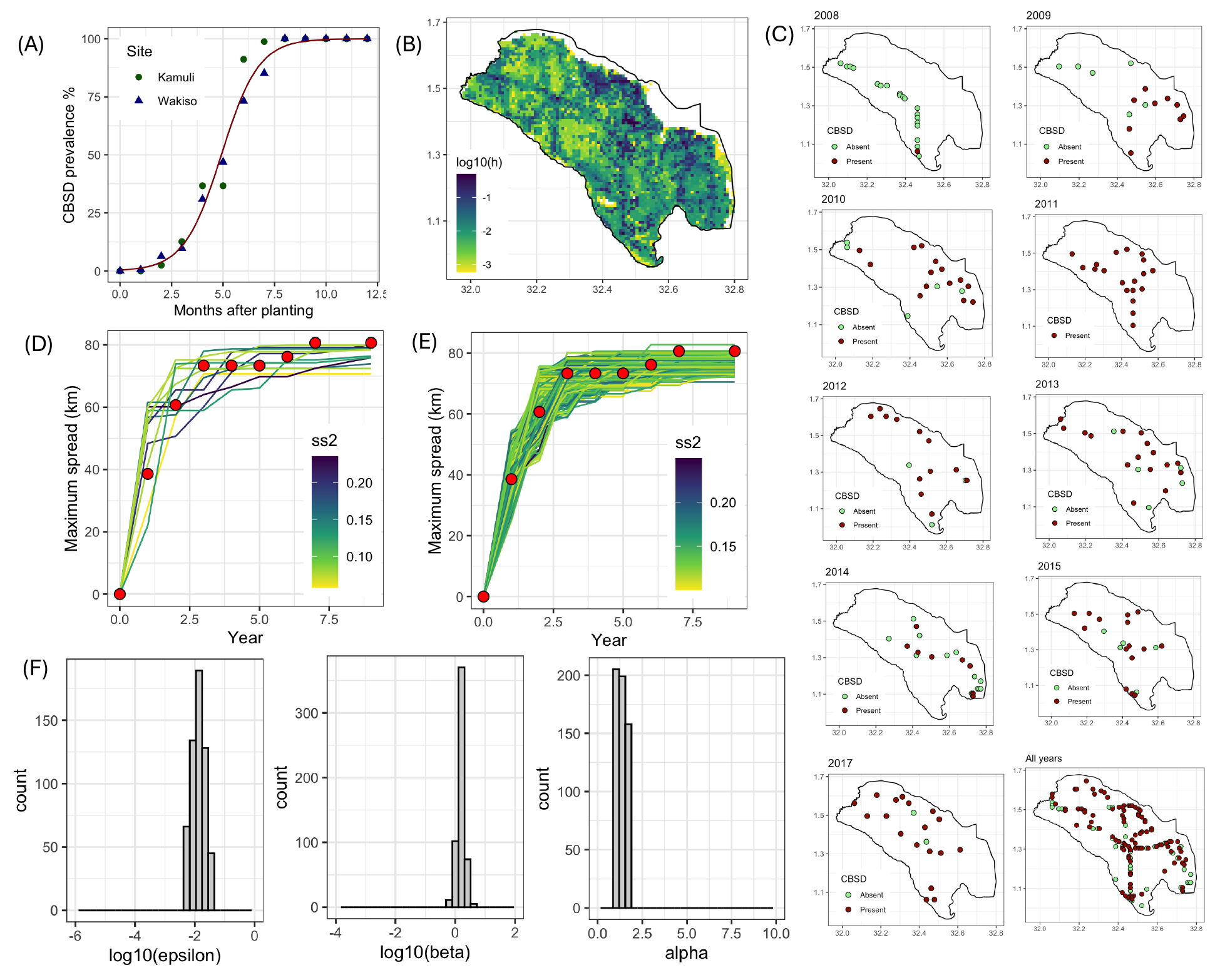}
\caption{(A) Progression of the fraction of infected hosts over time within cell: data from \citep{Katono2015} (blue and green points); and fitted logistic equation (red line). (B) The model host landscape, representing the fraction of 1 $km^2$ grid cell occupied by cassava, derived from CassavaMap \citep{Szyniszewska2020}. (C) Surveillance data from 2008 to 2017 for Nakasongola district \citep{Alicai2019}. (D) Summary statistics of particles accepted after stage 1. (E) Summary statistics of particles accepted after stage 2. (F) Marginal posterior distributions of parameters. }\label{figCase3}
\end{figure}

We set the prior distributions as follows: $\log10(epsilon)\sim U[-6,0]$, $\log10(\beta)\sim U[-4,2]$,and $\alpha\sim U[0.01,50]$. In the first stage of the ABC-RF rejection algorithm, we sampled $10^6$ parameter sets from the prior distributions and simulated the model. After  stage 1, there were 12 accepted parameter sets (Fig.\ref{figCase3}D). 

In the second stage, parameter sets that met the acceptance criteria were labeled as “accepted” (class 1) and used to train a Random Forest classifier. Following this, $10^7$ parameter sets were drawn from the prior distributions, and the trained classifier was applied to estimate their probability of acceptance. Parameter sets with predicted acceptance probabilities greater than 0.5 were retained for further analysis. After simulations, 562 out of 1205 parameter sets satisfied both summary statistic criteria and formed the empirical posterior distribution (Fig.\ref{figCase3}E). Marginal posterior distributions are shown in Figure \ref{figCase3}F.

\section{Discussion}

The ABC-RF-rejection algorithm introduced in this study marks a significant methodological advancement for parameter inference in complex epidemiological models, especially those that are both stochastic and spatially structured. By combining Approximate Bayesian Computation (ABC) rejection sampling with Random Forest classification, this two-stage framework improves computational efficiency while maintaining high inferential accuracy. Our results show that the ABC-RF-rejection approach substantially reduces the number of required simulations to approximate posterior distributions compared with conventional ABC rejection methods. Defining efficiency as the ratio of parameter samples contributing to the posterior distribution relative to the total number of simulations required across both stages, the algorithm achieved efficiencies of 0.19, 0.06, and 0.05 across the three case studies. This represents a considerable improvement over the traditional ABC rejection algorithm, which demonstrated an efficiency of just 0.0035 for parameter estimation in the  deterministic SIR model in Case 1, and efficiency of 0.006 for a stochastic spatial model of CBSD spread throughout Uganda \citep{Godding2023}. The observed gains in efficiency are attributed to the Random Forest classifier’s capacity to capture complex, nonlinear relationships between model parameters and the likelihood of producing simulations consistent with observed data.

The algorithm also demonstrates flexibility in handling multiple and potentially high-dimensional summary statistics, which are often necessary when dealing with heterogeneous data sources or spatially explicit models \citep{Beaumont2010}. In the application to a stochastic, spatially structured epidemic models (case 2), the framework successfully incorporated distinct summary statistics that captured radial disease spread and infection intensity, reflecting the multifaceted nature of epidemic dynamics. Notably, the algorithm facilitates the use of novel, context-specific summary statistics designed for settings where comprehensive surveillance data are scarce or incomplete \citep{Marjoram2003}.

While the ABC-RF-rejection approach offers clear advantages, certain limitations warrant consideration. The initial stage, which involves generating a large, labelled dataset of accepted and rejected parameter sets via ABC rejection, remains computationally intensive. This step is necessary to train the Random Forest classifier effectively, as its predictive performance hinges on the quality and diversity of the training data. This challenge is not unique to ABC-RF-rejection but is a common feature of many simulation-based inference algorithms \citep{Retkute2021}, where the stability and reliability of the results depend on sufficiently comprehensive exploration of the parameter space during the initial iteration.

Furthermore, the selection of the acceptance probability threshold in the second stage, which determines which parameter sets are retained for posterior approximation, introduces an element of subjectivity. This threshold must be chosen carefully, as overly stringent criteria may exclude plausible parameter sets, while overly permissive thresholds will increase computational cost. This trade-off is further complicated by the presence of stochasticity within many models, where variability in simulation outcomes arises even for identical parameter sets. Such stochasticity can blur the distinction between accepted and rejected simulations, making it more difficult to define a sharp threshold for acceptance. As a result, careful calibration of the acceptance probability threshold is essential \citep{Lintusaari2016}, particularly for complex or stochastic models where simulation outcomes exhibit high variability across runs.

Another consideration relates to the scalability of the approach for models with higher-dimensional parameter spaces. Although the current study focuses on models with relatively low parameter dimensionality, the capacity of the Random Forest classifier to maintain predictive accuracy in higher-dimensional settings remains to be fully explored. Nevertheless, given the algorithm’s demonstrated performance across both deterministic and stochastic models, there is considerable potential for its extension to more complex epidemiological systems, including those with greater parameter complexity or hierarchical structures.

Finally, the results highlight the potential for real-time or iterative parameter updating as new data become available, an increasingly important feature for informing disease management under evolving epidemic conditions. While formal comparisons to other methods were beyond the scope of this study, the distinct architecture of ABC-RF-rejection offers a promising alternative for situations where rapid, data-driven parameter inference is critical.

Overall, ABC-RF-rejection provides a flexible and efficient framework for parameter estimation in complex epidemiological models, with particular relevance for scenarios characterised by limited data availability, spatial heterogeneity, and the need for timely decision support.

\subsubsection*{Funding}
This research was funded by the Bill and Melinda Gates Foundation grant INV070408. The funder had no role in study design, data collection and analysis, decision to publish, or preparation of the manuscript.

\bibliographystyle{elsarticle-harv} 
\bibliography{References}

\end{document}